# On a new method to analyse QSO spectra

*Pierre Darriulat*
*Hanoi University of Sciences and VATLY Astrophysics Laboratory*
*INST, 179 Hoang Quoc Viet, Cau Giay, Ha Noi, Vietnam*

A new method of analysis of QSO spectra, usually referred to as the "Thong method", has been recently presented and made use of in a number of publications [1,2]. Several of these have been withdrawn because the authors have been convicted of plagiarism. However, there exists no publication showing that the method itself, which is an original contribution of the authors, is wrong. The purpose of the present note is to show that it is and that the results obtained when using it, including limits on the time variation of the fine structure constant many times smaller than published by other authors, must therefore be ignored and discarded.

To a good approximation, the relative splitting between the wavelengths of the lines of a spin-orbit doublet, $\Delta\lambda/\lambda$, is red-shift free and proportional to the square of the fine-structure constant $\alpha$. The light detected today, at time 0, from a high red-shift quasar was emitted at a time $t<<0$. Comparing the associated value of $\Delta\lambda(t)/\lambda(t)$ with that measured today in the laboratory, $\Delta\lambda(0)/\lambda(0)$ provides therefore a measure of a possible variation $\Delta\alpha$ of $\alpha$ between $t$ and now:

$$1+2\Delta\alpha/\alpha= [\Delta\lambda(t)/\lambda(t)]/[\Delta\lambda(0)/\lambda(0)] \qquad (1)$$

This is essentially Relation (1) of Reference 1. It is implicitly assumed that $\Delta\lambda<<\lambda$ and that $\Delta\alpha<<\alpha$, which indeed corresponds to reality. Exchanging the members of the doublet, which means changing the sign of $\Delta\lambda$, leaves Relation (1) invariant as expected.

We now review the seven relations of Reference 1 which summarize the "Thong method" and point to the mistakes which they contain. The seven relations are labelled (3) to (9) and are accompanied by very little explanatory text. Some of the mistakes could be interpreted as misprints, but they appear in several publications. Moreover, correcting such supposed misprints does not help with the understanding of the method. We therefore refrain from suggesting how the errors should or might be fixed.

Relation (3) writes the *"energy of a fine-structure level"* as:

$$E_{L,S,J}=E_0+\tfrac{1}{2}A(\alpha Z)^2+[J(J+1)-L(L+1)-S(S+1)]+B_J(\alpha Z)^4+\ldots \qquad (3)$$

Assuming that one uses natural units ($\hbar=c=1$), $A$ and $B_J$ are seen to have dimensions of energy while the term in square bracket is a pure number (corresponding to $2\mathbf{L.S}$). Relation (3) is therefore obviously wrong.

Relation (4) writes the *"relativistic corrections to the energy level transitions of atoms in the laboratory"* as

$$E_{L,S,J}-E_{L,S,J-1}=1+(B_J-B_{J-1}) (\alpha(0)Z)^2/(AJ) \qquad (4)$$

Here again, the left hand side has dimension of energy and the right hand side is a pure number. Moreover, when $\alpha(0)Z \to 0$, $E_{L,S,J} - E_{L,S,J-1} \to 1$ (in which units?) instead of 0 as expected.

Relation (5) writes explicitly Relation (4) in the case of two transitions

$$E_{\lambda2}(0) - E_{\lambda1}(0) = \tfrac{1}{2}(B_2 - 3B_1 + 2B_0)\,(\alpha(0)Z)^2/A \tag{5}$$

Here again, the left-hand side is an energy and the right-hand side a pure number.

Relation (6) uses Relation (3) to obtain:

$$\lambda_2(0)/\lambda_1(0) = 2 + (B_2 - 3B_1 + 2B_0)\,(\alpha(0)Z)^2/A \tag{6}$$

While dimensionally correct, this relation implies that when $\alpha(0)Z \to 0$, the wavelength ratio $\lambda_2(0)/\lambda_1(0)$ tends toward 2 instead of 1. This reveals a clear asymmetry between the roles played by the members of the doublet, demonstrating that Relation (6) is obviously wrong.

Relation (7) results from a comparison between Relations (5) and (6) and "*leads to the final result*":

$$E_{\lambda2}(0) - E_{\lambda1}(0) = \tfrac{1}{2}\lambda_2(0)/\lambda_1(0) - 1 \tag{7}$$

where again an energy is set equal to a pure number. Moreover when $\lambda_2(0) = \lambda_1(0)$ one obtains $E_{\lambda2(0)} - E_{\lambda1(0)} = -1/2$ (in which units?), which is surprising.

Relation (8) extends Relation (7) to time $t$

$$E_{\lambda2}(t) - E_{\lambda1}(t) = \tfrac{1}{2}\lambda_2(t)/\lambda_1(t) - 1 \tag{8}$$

and the comments made on Relation (7) apply equally here.

Finally, Relation (9) gives "*the width separation ratio between two lines from quasars and laboratory*". It summarizes the "Thong method" and has been used in several publications to obtain limits on a possible variation of the fine structure constant with cosmic times. It reads:

$$\alpha^2(t) = \alpha^2(0)\{\tfrac{1}{2}\lambda_2(t)/\lambda_1(t) - 1\}/\{\tfrac{1}{2}\lambda_2(0)/\lambda_1(0) - 1\} \tag{9}$$

which is not invariant in the exchange of $\lambda_1$ with $\lambda_2$ as it obviously should be if it were correct.

As Relation (9) is of the form $\alpha^2(t) = \alpha^2(0)F(t)/F(0)$, it implies $\alpha^2(t) = \alpha^2(0)$ whenever $F(t) = F(0)$, however arbitrary $F$ might be. As it is indeed the case, it is not surprising that the "Thong method" obtains small values [1,2] of $\alpha^2(t) - \alpha^2(0)$. However, these values are not correct and must be discarded.

It is not the place here to comment further on the "Thong method", it should be sufficient to have made it clear that it is not only wrong but also meaningless and that the results obtained using it should be ignored. None of the relations on which it is based is free of error.

References
1 Thong Duc Le, Lett. Prog. Theor. Phys. 126/1 (2011) 177.
2 Le Duc Thong and Nguyen Thi Thu Huong, Natural Science 3/7 (2011) 508.

L.D. Thong, N.M. Giao, N.T. Hung and T.V. Hung, Eur. Phys. Lett. 87 (2009) 6902. This article was formally withdrawn on 28 June 2010 by the Executive Editor *"on ethical grounds"*.

L. D. Thong *et al.*, Astrophysics, 53/3 (2010) 446.

L. D. Thong *et al.*, Astrophysics and Space Science, DOI 10.1007/s10509-010-0431-x (2010).

L.D. Thong and N. T. T. Huong (2011), "Studying cosmological time variability of the fine-structure constant from the analysis of quasar spectra", to appear in Journal of Modern Physics.